\documentstyle[11pt,IAUS215,twoside]{article}

\begin{document}
\vspace*{1cm}

\title{Absolute Wavelength Shifts\\
-- A new diagnostic for rapidly rotating stars}

\author{Dainis Dravins}

\affil{Lund Observatory, Box 43, SE-22100 Lund, Sweden}

\begin{abstract}  Accuracies reached in space astrometry now permit the accurate determination of {\it astrometric radial velocities}, without any use of spectroscopy.  Knowing this true stellar motion, spectral shifts intrinsic to stellar atmospheres can be identified, for instance gravitational redshifts and those caused by velocity fields on stellar surfaces.  The astrometric accuracy is independent of any spectral complexity, such as the smeared-out line profiles of rapidly rotating stars.  Besides a better determination of stellar velocities, this permits more precise studies of atmospheric dynamics, such as possible modifications of stellar surface convection (granulation) by rotation-induced forces, as well as a potential for observing meridional flows across stellar surfaces.
\end{abstract}

\section{Introduction}

Much of what we know about rotating stars is deduced from their spectra.  However, with increased rotational velocity, spectral lines become smeared-out, making the measurement and interpretation of stellar spectral features more difficult.

Recent developments in astrometry now permit the accurate determination of stellar radial motion without involving spectroscopy or the Doppler principle.  Although astrometry has normally not been associated with studies of stellar motion in the radial direction, astrometric quantities (e.g., proper motions) are sensitive to also the radial-velocity component (although only as a second-order effect).  The accuracies recently realized in space astrometry have made it practical to determine such {\it astrometric radial velocities} (Dravins, Lindegren, \& Madsen 1999; Madsen, Dravins, \& Lindegren 2002).  The differences between these values (giving motions of the stellar centers of mass with even sub-km~s$^{-1}$ accuracies) and apparent spectroscopic velocities reveal wavelength shifts intrinsic to stellar atmospheres, such as gravitational redshifts or those caused by gas flows on stellar surfaces.

With current techniques, the most accurate results are obtained for stars in open clusters, moving through space with essentially the same velocity vector.  Parallaxes give the distances, while proper-motion vectors show the fractional change with time of the cluster's angular size.  The latter equals the time derivative of distance, yielding the radial velocity.  Such young open clusters contain a significant number of rapidly rotating stars (Fig.\ 1).

\begin{figure}
\plotfiddle{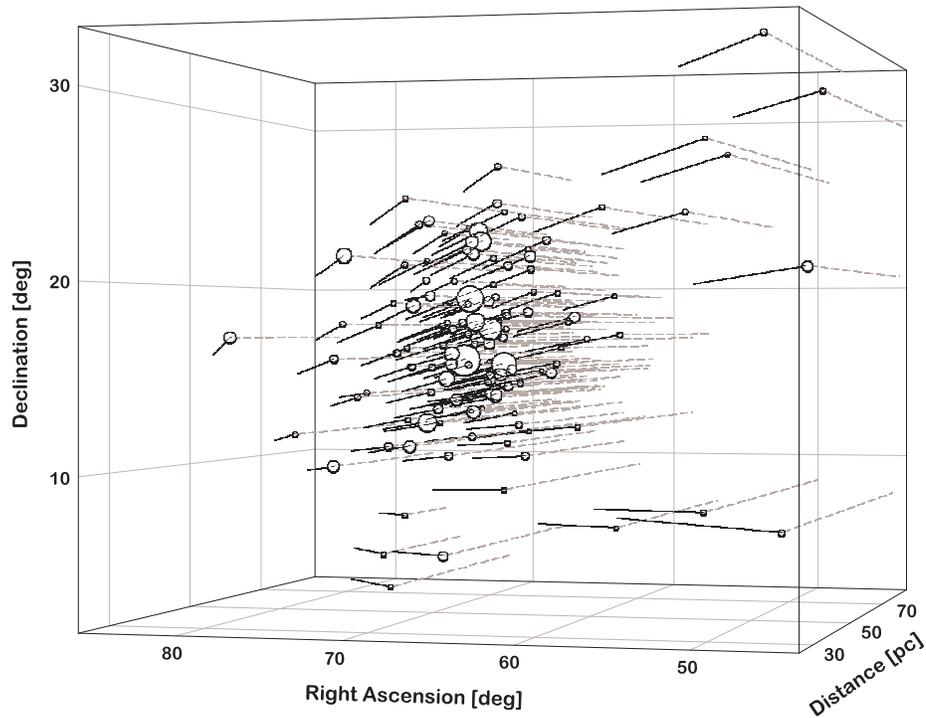}{10 cm}{0}{110}{110}{-260}{0}
\caption{One principle for the determination of astrometric radial velocities: Proper motions in the Hyades as measured by Hipparcos, with symbol size indicating stellar brightness.  Stars in a moving cluster share the same average velocity vector.  Parallaxes give the distance, while proper-motion vectors show the fractional change with time of the cluster's angular size.  The latter corresponds to the time derivative of distance, yielding the radial velocity (Dravins et al.\ 1997)}
\end{figure}

\section{Lineshifts intrinsic to stellar atmospheres}

On Hipparcos, an observing program was carried out for stars in moving clusters, yielding astrometric radial-velocity solutions for more than 1,000 stars (Madsen et al.\ 2002), including about one hundred stars in the Hyades and Ursa Major groups.  The error budgets are somewhat complex, but accuracies can reach around 0.5 km~s$^{-1}$ (Madsen 2003).

Differences between astrometrically determined radial velocities and the apparent spectroscopic ones reveal lineshifts intrinsic to stellar atmospheres.  Fig.~2 indicates that F-type spectra have enhanced blueshifts relative to cooler ones (as actually expected from hydrodynamic simulations), a trend that seems to continue to even hotter stars (where corresponding hydrodynamic models are not yet developed).

The likely origin of such wavelength shifts can be traced back to stellar surface structure.  While this can be studied in detail on the Sun, corresponding phenomena in other stars are mainly seen as subtle asymmetries in photospheric spectral lines.  Bright (hot) elements in solar (and stellar) granulation are rising, causing local blueshifts.  A spectral line averaged over the stellar disk then obtains a {\it convective blueshift} since the bright (and blueshifted) elements statistically contribute a greater number of photons.  Three-dimensional hydrodynamic models reproduce both the granulation patterns and the resulting wavelength shifts (Dravins \& Nordlund 1990; Allende Prieto et al.\ 2002).

\begin{figure}
\plotfiddle{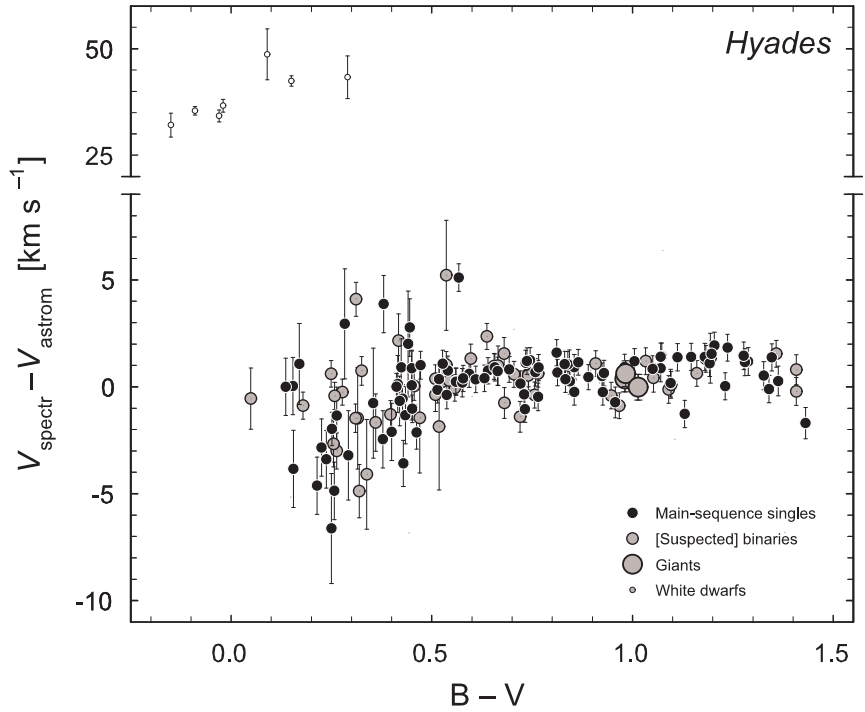}{9.5 cm}{0}{120}{120}{-175}{-210}
\caption{ The Hyades: Differences between spectroscopic radial-velocity values from the literature, and astrometric determinations.  An increased blueshift of spectral lines in stars somewhat hotter than the Sun ($\bv \simeq 0.3-0.5$) is theoretically expected due to their more vigorous surface convection, causing greater convective blueshifts.  Gravitational redshifts of white-dwarf spectra place them far off main-sequence stars.  The error bars show the combined spectroscopic and astrometric errors (Madsen et al.\ 2002)}
\end{figure}

\subsection{ Effects for rotating stars }

Increased stellar rotation produces an increased centrifugal force and lowers the effective surface gravity.  Analogous to modeled differences between hydrodynamic atmospheric models for dwarfs and lower-gravity subgiants, more vigorous granulation might then be expected to develop (since the convective energy flux must then be carried by a lower-density gas).  The greater velocity amplitudes and temperature contrasts become visible in integrated starlight as enhanced convective blueshifts.  Further effects could perhaps arise from Coriolis forces, which might influence the granulation structure in rapidly rotating stars.

A tantalizing possibility is the potential of observing {\it meridional flows} across stellar surfaces.  A circulation pattern between the equator and poles will involve rising (or sinking) gases at the equator, with the opposite flows near the poles.  Stars seen equator-on (with a large observed value of Vsin$i$) will then have their spectra influenced by Doppler shifts from the systematic upflow (or downflow) patterns there, while stars seen pole-on (with a small value of Vsin$i$) will show lineshift signatures of the opposite sign.

Since various effects of atmospheric dynamics will combine, a segregation of the various contributions will require a detailed physical modeling, in particular exploiting the fact that the exact wavelength displacements depend on the properties of various spectral lines.  For example, convective blueshifts differ among lines of different strength (formed at different depths), of different excitation potential and ionization level (formed preferentially in differently hot elements), and of wavelength region (with different intensity contrasts between the hotter and cooler elements).

For stars in the Hyades, a Vsin$i$ dependence of the intrinsic wavelength shift is observed -- spectra of rapidly rotating stars appear blueshifted relative to those of slower rotators (Fig.\ 3).  While such a dependence is thus perhaps not unexpected, any interpretations must still be viewed with caution since this dependence also correlates with that on temperature (spectra of earlier-type stars are blueshifted relative to those from cooler ones; Fig.\ 2), and rapid rotation statistically dominates for earlier types.

\begin{figure}

\plotfiddle{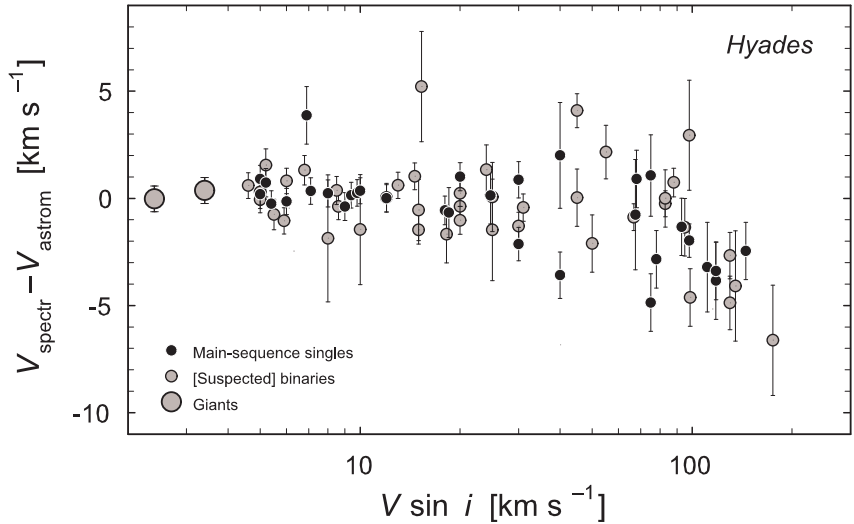}{9 cm}{0}{120}{120}{-170}{0}
\caption{ The Hyades: Differences between spectroscopic and astrometric velocities as in Fig.\ 2, as function of stellar rotational velocity Vsin$i$.  Original figure from Madsen et al.\ 2002, here updated with additional rotational data from G\l\c{e}bocki \& Stawikowski (2000.)}
\end{figure}

\subsubsection{Spectroscopic effects of stellar rotation}

The accurate determination of spectroscopic wavelength shifts in rotating stars can involve some complications since the precise value may depend on precisely how the wavelength shifts are measured.  For example, radial-velocity instruments often apply a cross correlation of the observed spectrum with a spectrum template taken from a slowly rotating star.  However, for a rapidly rotating early-type star, a typical template may have a mismatch exceeding 1~km~s$^{-1}$, due mainly to the rotational broadening and the ensuing blending of spectral lines (Verschueren \& David 1999; Griffin, David, \& Verschueren 2000).

Even modest rotational velocities in sharp-lined late-type stars may cause wavelength displacements of the spectral-line bottoms used for radial-velocity determinations.  For sufficiently rapid rotation, when asymmetric line components originating near the stellar limbs begin to affect the wings of the profile integrated over the stellar disk, the intrinsic line asymmetries may become enhanced (Gray \& Toner 1985; Gray 1986; Smith, Livingston, \& Huang 1987; Dravins \& Nordlund 1990).

A conclusion to be drawn is (contrary to a sometimes expressed belief) that spectroscopic observations of rapidly rotating stars need to be made not only at a high photometric signal-to-noise ratio, but also at very high spectral resolution -- much higher than that corresponding to the stellar rotational velocity itself.

\section{ A new diagnostic for rapidly rotating stars }

In the past, absolute lineshifts could be studied only for the Sun (since the relative Sun-Earth motion is known from planetary-system dynamics, and does not rely on spectral measurements).  For other stars, this is now becoming possible thanks to three separate developments: (a) Astrometric measurements permit to accurately determine stellar radial motion without using any spectroscopy; (b) The availability of high-resolution spectrometers with accurate wavelength calibration (such as designed for exoplanet searches), and (c) Accurate laboratory wavelengths for several atomic species.  These developments appear particularly significant for the study of stars with complex or smeared-out spectra, such as rapidly rotating ones, or such with large-scale atmospheric motions or mass loss.  For a further discussion, see Dravins (2003).

\acknowledgments

This work is supported by The Swedish Research Council, the Swedish National Space Board, and The Royal Physiographic Society in Lund.

\end{document}